# Excitation of forward and backward surface waves in a negative-index metamaterial

A. Isa[1],

[1] Department of Physics, Federal University Lokoja,
P.M.B 1154, Lokoja, Kogi State, Nigeria

Isa.ali@fulokoja.edu.ng

**Abstract**

We study beam transmission and refection resulting from the resonant excitation of forward and backward surface waves in a metamaterial. We predict that exciting these waves at an interface this medium and free space led to transmitted and reflected beams with high amplitudes. The FDTD simulated results demonstrate clearly propagation of these waves and their energy-trapping ability as described by theory. The resonant coupling of the refracted waves and reflected waves at the first interface led to the excitation of the backward surface waves. The forward surface waves transfer incoming energy in the direction of incidence and the backward surface transfer incoming energy in a direction opposite to the direction of incidence (i.e. negative direction). More also, the parameters used in these simulations were carefully chosen to minimize the effects of interference between the incident and the reflected beams earlier reported. In overall, these numerically simulated results clearly demonstrate the propagation of these waves and may serve as numerical approximations to what might be observed in experimental setups. Putting this simply, they may find applications in optical and photonic devices involving enhancing beam transmission and reflection using negative-index metamaterial.

1. Introduction

Theoretical [1] and experimental [2] studies have shown the propagation of surface waves at the interface of the negative index-material and free space for both TM and TE polarizations [3, 4]. These waves are capable of transferring the energy of the incident wave along the interface of the material medium. This leads to effective enhancement of the lateral shift in the reflected and transmitted beams [5]. Besides, they can also enhance the value of the giant Goos – Hanchen effect in the reflected beam. By the same token, they can be resonantly excited when the tangential component of the wave vector coincides with the propagation constant of a corresponding surface polaritons [5]. Vortex-like surface waves have been proposed in Ref. [16] which do not transport wave energy along the interface but are responsible for the image oscillations often encountered when focusing with a superlens [16].

Previous studies have also shown the existence of forward and backward propagating surface polaritons at the interface of the negative-index material [4]. For a three-layered structure, their excitations depend on the effective parameters of the media given by $X =|\varepsilon_3|/\varepsilon_2$ and $Y =|\mu_3|/\mu_2$. In this structure, the second layer was chosen to be air to allow observation surface waves when excited. The parameters $\varepsilon$ and $\mu$ are the permittivity and permeability respectively.

While forward propagating surface waves transfer wave energy in a direction of the incidence [5], the backward surface waves transfer energy in a direction opposite to the direction of incidence (i.e. negative direction) [5]. To excite the backward propagating surface waves, the following conditions were met using the effective parameters given by $XY >1$ and $Y <1$. This results in a negative shift in the reflected beam which can be observed in [5, Fig. 4(a)]. In an attempt to substantially minimize the effects of interference between the incident and reflected beams reported in [5, Fig. 4], we used a different approach to excite both the forward and backward propagating surface waves by carefully considering the simulation parameters in the FDTD algorithm: source-to-slab distance, slab thickness, source type and of course changing the refractive index of the material medium. The simulated FDTD results revealed that the backward propagating surface waves are resonantly excited when the refracted wave couples with a reflected wave at the first interface.

**FINITE DIFFERENCE TIME DOMAIN ALGORITHM**

The FDTD algorithm explores the electromagnetic problem for TM sets with $H_x, H_y$ and $E_z$. We discretized electric and magnetic fields in space and time. In other words, the fields are allocated in space and matching in time for the evolution of the procedure as described by Yee's robust numerical scheme [8]. This robust scheme considers the electric and magnetic fields being staggered in space by half a cell and in time by half a time step using their partial derivatives. The discretized update equations for the Lorentz model used in this simulation are given below and detailed information can be found in [9, 10]:

$$I_k^{n+1} = \left[\frac{2-\Delta t^2 \omega_{oe}^2}{1+0.5\Delta t \Gamma_e}\right]I_k^n + \left[\frac{0.5\Delta t \Gamma_e - 1}{0.5\Delta t \Gamma_e + 1}\right]I_k^{n-1} + \left[\frac{\Delta t^2 \varepsilon_0 \omega_{pe}^2}{1+0.5\Delta t \Gamma_e}\right]E_k^n \quad (1)$$

$$E_z^{n+1}(i,j) = \left(\frac{2\varepsilon - \Gamma_e \Delta t}{2\varepsilon + \Gamma_e \Delta t}\right)E_z^n(i,j) + \frac{2\Delta t}{2\varepsilon + \Gamma_e \Delta t}\left[\frac{H_y^n(i,j) - H_y^n(i-1,j)}{\Delta x}\right]$$
$$- \frac{2\Delta t}{2\varepsilon + \Gamma_e \Delta t}\left[\frac{H_x^n(i,j) - H_x^n(i,j-1)}{\Delta y}\right] \quad (2)$$

$$D_z^{n+1}(i,j) = D_z^n(i,j) + \frac{\Delta t}{\Delta x}\left(H_y(i,j) - H_y(i-1,j)\right) - \frac{\Delta t}{\Delta y}\left(H_x(i,j) - H_x(i,j-1)\right) \quad (3)$$

$$E_z^n(i,j) = \frac{(D_z^n - I_k^n)}{(\varepsilon_\infty \varepsilon_0)} \quad (4)$$

$$G_k^{n+1} = \left[\frac{2-\Delta t^2 \omega_{om}^2}{1+0.5\Delta t \Gamma_m}\right]G_k^n + \left[\frac{0.5\Delta t \Gamma_m - 1}{0.5\Delta t \Gamma_m + 1}\right]G_k^{n-1} + \left[\frac{\Delta t^2 \mu_0 \omega_{pm}^2}{1+0.5\Delta t \Gamma_m}\right]H_k^n \quad (5)$$

$$B_y^{n+1}(i,j) = B_y^n(i,j) + \frac{\Delta t}{\Delta x}\left(E_z^n(i+1,j) - E_z^n(i,j)\right) \quad (6)$$

$$B_x^{n+1}(i,j) = B_x^n(i,j) - \frac{\Delta t}{\Delta y}\left(E_z^n(i,j+1) - E_z^n(i,j)\right) \quad (7)$$

$$H_z^n(i,j) = \frac{(B_y^n - G_k^n)}{(\mu_\infty \mu_0)} \quad (8)$$

$$H_x^{n+1}(i,j) = \left(\frac{2\mu - \Delta t \Gamma_m}{2\mu + \Delta t \Gamma_m}\right)H_x^n(i,j) - \frac{2\Delta t}{2\mu + \Delta t \Gamma_m}\left[\frac{E_z^n(i,j) - E_z^n(i,j-1)}{\Delta y}\right] \quad (9)$$

$$H_y^{n+1}(i,j) = \left(\frac{2\mu - \Delta t \Gamma_m}{2\mu + \Delta t \Gamma_m}\right)H_y^n(i,j) - \frac{2\Delta t}{2\mu + \Delta t \Gamma_m}\left[\frac{E_z^n(i+1,j) - E_z^n(i,j)}{\Delta x}\right] \quad (10)$$

The following procedures were used to run the FDTD simulations using the update equations above.

(a) Update $I_k$ in Eq. (1) using the previous values of $E$ and $I_k$.

(b) Update $E$ in Eq. (2) in free space and add a source.

(c) Update $D$ in Eq. (3) in the material medium using the previous values of $D$ and $H$.

(d) Update $E$ in Eq. (4) using the previous values of $D$ and $I_k$.

(e) Apply the Mur's ABC at the four sides of the simulation domain.

(f) Repeat the above procedures for the magnetic fields using Eq. (5) – Eq. (10).

## 2. FDTD SIMULATION RESULTS AND DISCUSSIONS

We carried out a causal FDTD numerical simulation to study the dynamics of exciting forward and backward propagating surface waves at the interface of a negative index medium and free space. The FDTD simulation space was 400 cells in the x-direction and 400 cells in the y-direction. The simulation parameters were carefully chosen to minimize the effect of interference between the incident and the reflected beams which earlier reported [5]. We also attempt to evaluate the contribution of these surface waves to the transmitted and reflected beams at the interface of the model. The negative-index medium was modeled with a lossy Lorentz model for both permeability and permittivity and a time step of $\Delta t = 2.239 \times 10^{11} s$ was used in the simulations. Optimal cell size $\Delta x = \Delta y = \lambda/40$ [9, 10] was used to minimize the effects of numerical dispersion associated with the numerical schemes.

The FDTD grid was terminated by MUR's absorbing boundary conditions on all four sides of the simulation domain [12]. A source was positioned at a distance $y_0 = 0.5\lambda$ from the slab with thickness $d = 2\lambda$. The negative-index material was matched to free space such that $w_{0e} = w_{0m} = w_0$, $w_{pe} = w_{pm} = w_p$ and $\Gamma_e = \Gamma_m = \Gamma$. The center frequency of the Gaussian beam source was 6 GHz which corresponds to $\lambda_0$ = 5cm and this was driven by a continuous wave pulse [13].

A low loss value of $\Gamma = 1 \times 10^7 \, rad/s$ was used throughout the simulations and for a medium with refractive index $n(\omega) \approx -1$, the following parameters were used $w_p = 6.9285 \times 10^9 \, rad/s$, $w_0 = 9.9965 \times 10^8 \, rad/s$. For the case of material medium with $n(\omega) \approx -2$, the plasma and resonance angular frequencies are given as $w_p = 6.9285 \times 10^9 \, rad/s$, $w_0 = 9.4185 \times 10^8 \, rad/s$.

For $n(\omega) \approx -6$, $w_p = 8.6318 \times 10^9 \, rad/s$, $w_0 = 1.8836 \times 10^9 \, rad/s$ [9, 10].

A sinusoidal Gaussian beam was driven into the metamaterial medium with the source having cell distance $\Delta$ and these cells are excited by a sinusoidal source $\sin(2\pi f_0 t)$. There are 25 line sources that were formed with a distance of $5\Delta$. The incident angle of the beam was kept at $\theta_{inc}$ = 25° to enable excitation of surface waves and a weighting function $\exp\left(\dfrac{-x^2}{nT^2}\right)$ was also used, where $nT$ controls the beam width of the sinusoidal source [11]. The electric field distribution over the FDTD simulation domain for different refractive indices are shown in Figs. 1(a, b) and Figs. 2(a, b).

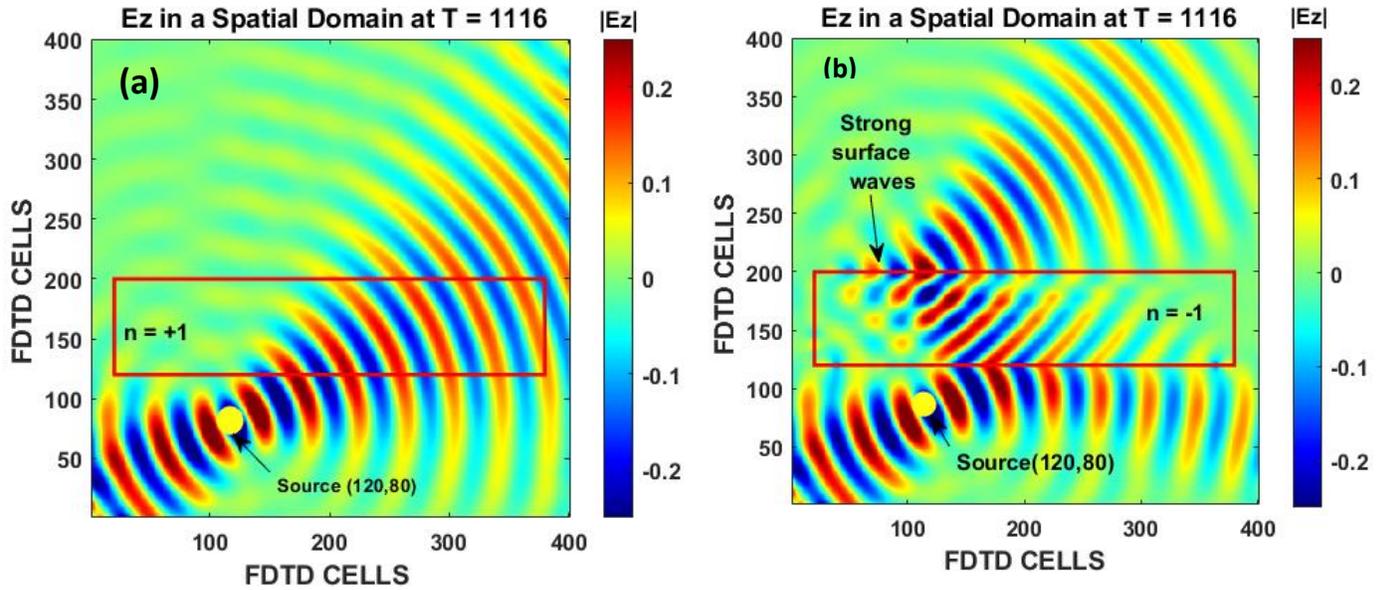

Fig. 1: The electric field distribution over the FDTD domain (a) for a positive-index medium at time step t = 1116 $\Delta t$. (b) For a negative-index medium with $n(\omega) \approx -1$ at time step t = 1116 $\Delta t$.

Comparing Fig. 1 (a) and Fig. 1(b), revealed the negative refraction exhibited by the metamaterial with a refractive index $n(\omega) \approx -1$. As reported in [5], forward and backward propagating surfaces waves are generated when the beam interacts with the metamaterial and these can be observed in Fig. 1(b) and Fig. 2(a) respectively. Forward propagating surface waves are visibly present in Fig. 1(b). They are excited when a refractive wave couples with a transmitted wave at the second interface. The intensity of the transmitted beams is stronger when compared to those in Figs. 2(a, b). This can be attributed excited forward surface waves which enhanced the transmitted beams in Fig. 1(b). This simulated result clearly demonstrated the capability of these waves transferring along the interface as earlier predicted [5].

Additionally, the backward surface waves propagate along the first interface in a direction opposite to the direction of incidence and thus transfer wave energy that direction (Fig. 2(a)). They are excited when there is a resonant coupling of a refracted wave and a reflected wave at the first interface (Fig. 2(a)). The physical consequence of exciting these surface waves is that the incident beams are strongly reflected as demonstrated in Fig. 2(a).

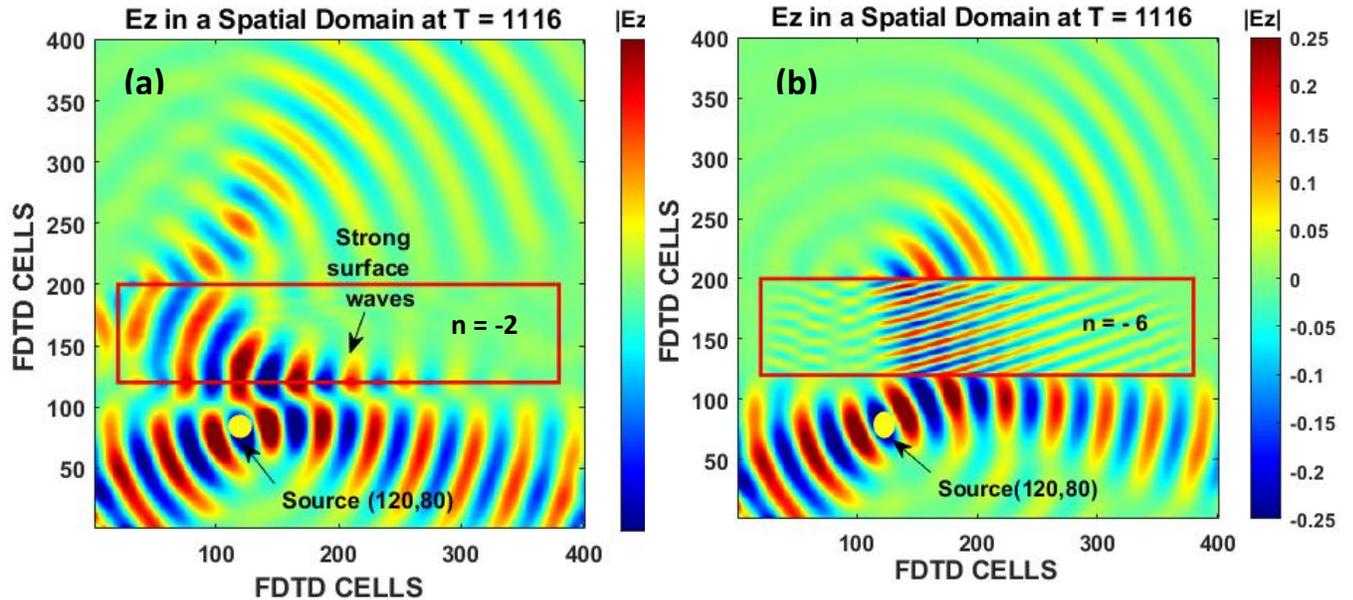

Fig. 2: The electric field distribution over the FDTD domain (a) for negative-index material with $n(\omega) \approx -2$ t = 1116 $\Delta t$ (b) for a negative-index material with $n \approx -6$ at t = 1116 $\Delta t$.

In contrast to the simulated results for models with refractive indices $n(\omega) \approx -1$ and $n(\omega) = -2$, there are no visible surface waves generated in the model with refractive index $n(\omega) \approx -6$. More also, the angles of refraction in this model are shallower when compared with those in Fig. 1(b) and Fig. 2(a). However, a strong beam compression occurred along the center of the model which may be attributed to the decrease in the wavelength of the incident beam as the refractive index increases by a factor of 6 [6, 7].

In summary, we performed a causal finite difference time domain simulations with a twin goals of exciting the forward and backward surface waves and of course evaluating their energy trapping mechanisms. Visual inspection revealed strong intensity for the transmitted beams through the model with refractive index $n(\omega) \approx -1$ and this was attributed to the excitation of the forward surface waves along second interface. No visible surface waves were excited in the model with refractive index of n $n(\omega) \approx -6$ but however axial beam compression was observed. Besides, the incident beam was strongly reflected for the medium with refractive index $n(\omega) \approx -6$ and this was due to the excitation of backward surface waves.

These numerically simulated results may serve as useful approximations when considering the excitation of both surface and backward surface waves in experimental setup. They also demonstrate the choice of refractive index to be used when exciting these surface waves. In overall, these results may find applications in beam propagation through optical and photonic devices with negative refractive indices.